\documentclass[hyper,letterpaper,12pt]{article}
\usepackage{a4wide}
\usepackage{amsmath}
\usepackage[
      colorlinks=true,
      linkcolor=blue,
      urlcolor=blue,
      filecolor=black,
      citecolor=blue,
            ]{hyperref}
\def\nn{{\nonumber}}
\def\beq{\begin{equation}}
\def\eeq{\end{equation}}
\def\p{\partial}

\def\b{{\beta}}
\def\a{{\alpha}}
\def\g{{ \gamma}}
\def\d{{\delta}}
\def\t{{ \theta}}
\def\e{{\epsilon}}

\def\c{\theta}
\def\sc{$\Sigma_c$}
\def\x{\tilde{x}}

\def\r{\rho}
\def\o{\omega}
\def\R{\hat R}
\def\R{\hat R}

\def\u{\tilde{u}}

\def\T{\tilde{T}}
\def\h{\tilde{h}}

\def\si{\tilde{\sigma}}
\def\pa{\tilde{\partial}}
\def\t{\tau}
\def\tc{\tilde{\tau}}
\def\m{\mu}
\def\n{\nu}

\renewcommand{\(}{\left(}
\renewcommand{\)}{\right)}
\renewcommand{\[}{\left[}
\renewcommand{\]}{\right]}

\begin{document}

\title{\bf \Large Non-Relativistic Fluid Dual to Asymptotically\\ AdS Gravity at Finite Cutoff Surface}

\author{\large
~Rong-Gen Cai\footnote{E-mail: cairg@itp.ac.cn}~,
~~Li Li\footnote{E-mail: liliphy@itp.ac.cn}~,
~~Yun-Long Zhang\footnote{E-mail: zhangyl@itp.ac.cn}\\
\\
\small Key Laboratory of Frontiers in Theoretical Physics,\\
\small Institute of Theoretical Physics, Chinese Academy of Sciences,\\
\small P.O. Box 2735, Beijing 100190, China\\}
\date{\small May 10, 2011}
\maketitle

\begin{abstract}
\normalsize Using the non-relativistic hydrodynamic limit, we solve equations of motion for Einstein gravity
and Gauss-Bonnet gravity with a negative cosmological constant within the region between a finite cutoff
surface and a black brane horizon, up to second order of the non-relativistic hydrodynamic expansion parameter.
Through the Brown-York tensor, we calculate the stress energy tensor of dual fluids living on the cutoff surface.
 With the black brane solutions,  we show that for both Einstein gravity and Gauss-Bonnet gravity, the  ratio of
  shear viscosity to entropy density of dual fluid does not run with  the cutoff surface.
  The incompressible Navier-Stokes equations are also obtained in both cases.
\end{abstract}

\section{ Introduction}
The AdS/CFT correspondence~
\cite{Maldacena:1997re,Gubser:1998bc,Witten:1998qj,Aharony:1999ti}
relates gravity in an anti-de Sitter (AdS) spacetime to a strongly coupled conformal field theory (CFT)
living on the boundary of the AdS space. Recently the AdS/CFT correspondence has been applied to various
fields. By use of the AdS/CFT correspondence, one can calculate many quantities of strongly coupled CFTs
through dual gravity theories.  A remarkable example is the calculation of the ratio of shear viscosity to
 entropy density $\eta/s$ of some field theories dual to the AdS Einstein
 gravity \cite{Policastro:2001yc,{Kovtun:2003wp},{Buchel:2003tz},{Kovtun:2004de}}.
In Einstein gravity, it was found that the ratio is a universal value $1/4\pi$, while in the case
with $R^2$ corrections \cite{Brigante:2007nu,{Brigante:2008gz},{KP},{Cai:2009zv}} there is a negative
 additional correction term.  Furthermore, it was shown that $\eta/s$ only depends on the value of
 effective coupling of transverse gravitons evaluated on the
horizon~\cite{Brustein:2008cg,Hong.Membrane,Cai:2008ph,Cai:2009zn}.
Under in the hydrodynamic limit, it was generally proven that some linear response coefficients are
universal  both in the AdS/CFT correspondence and membrane paradigm~\cite{Hong.Membrane}. In particular,
one can consider a fictitious membrane at a constant radial radius to express the AdS/CFT response in
terms of the membrane paradigm language~\cite{Hong.Membrane}.  The dependence of
the diffusion constant of dual fluid on the cutoff surface is interpreted as the Wilson renormalization
group flow~\cite{Strominger}. Some studies relating the radial radius of AdS space to energy scale of
dual CFT appear
in~\cite{Susskind,{Balasubramanian},de Boer:1999xf}. Some forms of holographic renormalizaton group flow
equation independent of the cutoff were given recently
in~\cite{Nickel:2010pr,Heemskerk:2010hk,Faulkner:2010jy} and it was generally proven in \cite{Sin:2011yh}
that they are actually equivalent to the radial evolution of the classical equation of motion \cite{Strominger}.

Since the Wilson  renormalization group flow theory does not require
an ultraviolet completion of quantum field theory, the authors of
\cite{Strominger} also do not insist on an asymptotically AdS
region, instead they introduce a finite cutoff $r_c$ outside the
horizon in a general class of $p+2$-dimensional black hole
geometries. The dispersion relation of the gravitational
fluctuations confined inside the cutoff is shown at long wavelengths
to be that of a linearized $p+1$-dimensional Navier-Stokes (NS)
fluid living on the cutoff
surface~\cite{Strominger,Bredberg,Skenderis}. This remarkable
relation was investigated in~\cite{Bredberg} that a given solution
of the incompressible NS equations maps to a unique solution of the
vacuum Einstein equations. An algorithm was presented in
\cite{Skenderis} for systematically reconstructing a solution for
the $p+2$-dimensional vacuum Einstein equations from a
$p+1$-dimensional fluid, to arbitrary order by extending the
non-relativistic hydrodynamic expansion proposed in~\cite{Bredberg}.

Clearly it is of great interest to develop a holography fluid
description dual to an asymptotically flat gravitational
configuration by introducing a finite cutoff. In this paper,
however, we discuss this issue in asymptotically AdS gravity by
introducing a finite radial cutoff, because when one takes the
cutoff to be infinity, the dual field theory on the AdS boundary is
well-defined and some results are comparable to those in the
literatures. Clearly such a study could be a service to further
discuss the case in asymptotically flat spacetimes.

Note that one can construct the stress energy tensor of the dual fluid order by order from the bulk gravity
solution~\cite{Bhattacharyya:2008jc}. In this paper we follow the procedures
 in \cite{Hong.Membrane,Strominger,Bredberg,Skenderis}, by
introducing a finite cutoff surface $\Sigma_c$ outside black hole horizon, generally discuss
the effect of finite perturbations of the extrinsic curvature of \sc\ while keeping the intrinsic
 metric of the cutoff surface flat \cite{Bredberg}. By applying two finite diffeomorphism transformations,
 in the non-relativistic hydrodynamic
expansion limit we obtain black brane solutions, up to second order, of the non-relativistic hydrodynamic expansion
parameter $\epsilon$, between the cutoff surface and the horizon in
Einstein gravity with a negative cosmological constant. We calculate
the stress energy tensor of the
 fluid on the cutoff surface. The results show that the ratio $\eta/s$ is still $1/4\pi$, independent
 of the cutoff, which implies that it does not run along the radial coordinate. And it turns out that
 the stress energy tensor of the fluid obeys the incompressible Navier-Stokes equations. We also
 discuss the case of Gauss-Bonnet gravity with a negative cosmological constant.

The paper is organized as follows. In Sec.\eqref{sect:tm} we
introduce two finite diffeomorphism transformations to a general metric dual to fluid in flat spacetime,
while keeping the induced metric of the cutoff surface invariant.  We make the non-relativistic hydrodynamic
expansion and solve gravitational equations to the second order of the expansion parameter.
In Sec.\eqref{sect:ein} we apply this formulism to Einstein gravity with a negative cosmological constant.
We find that the ratio of shear viscosity to entropy density of dual fluid on the finite cutoff
surface $\eta/s=1/4\pi$, independent of the cutoff and that the conservation equation of the stress energy
 tensor of the dual fluid gives an incompressible Navier-Stokes equation.  In Sec.\eqref{sect:gb}
 we consider the case of Gauss-Bonnet gravity with a negative cosmological constant. The ratio of
 shear viscosity to entropy density is found to be $\eta/s=(1-8\a)/4\pi$, and corresponding
 incompressible Navier-Stokes equations are also obtained there.  The
 conclusions and some discussions are included in Sec.\eqref{sect:co}.


\section{Non-relativistic hydrodynamic expansion}
\label{sect:tm}

To study the dynamics of fluid in $p+1$-dimensional
flat spacetime, we consider a generic $p+2$-dimensional metric:
\beq
ds_{p+2}^{2}=-h(r)d\tau^{2}+2d\tau
dr+a(r)dx_{i}dx^{i},\label{gm}
\eeq
where $h(r)$ and $a(r)$ are two functions of radial coordinate $r$. Introducing a finite
cutoff surface \sc ~at $r=r_c$ (outside black hole horizon if the horizon
is present), the induced metric on the surface is flat
\beq
\g_{ab}dx^a dx^b=-h(r_c)d\tau^2+a(r_c)dx_idx^i,
\eeq
where $x^a \sim (\tau,x^i)$. Introduce proper intrinsic coordinates $\x^a \sim (\tc,\x^i)$ on $\Sigma_c$ as
\beq
\x^0\equiv\tc=\sqrt{h(r_c)}\t ,~~~\x^i=\sqrt{a(r_c)}x^i\label{ct}
\eeq
the induced metric is simply given by
\beq ds_{p+1}^2=\eta_{ab}d\x^ad\x^b=-d\tc^2+\d_{ij}d\x^id\x^j.
\eeq
In order to keep the intrinsic metric of  \sc\  flat, following
\cite{Skenderis} we take two finite diffeomorphism transformations. The first one is a Lorentz boost
with a constant boost parameter $\beta_i$. In the $(\tc, ~\x^i)$ coordinates, it is given by
\beq
 \tc \rightarrow \g \tc-\g\beta_i \x^i, \qquad \x^i \rightarrow \x^i -\g\beta^i
\tc+(\g-1)\frac{\beta^i\beta_j}{\beta^2}\x^j,\label{t1}
\eeq
where
\beq \u^{a}=\g(1, ~\beta^i),~~~~ \g=\frac{1}{\sqrt{1-\beta^2}},~~~
\beta_i=\d_{ij}\beta^j \eeq
 In the $(\tau, ~x^i)$ coordinates, we have
\beq
u^a = \frac{\(1,~v^i\)}{\sqrt{h(r_c)-v^2}},~~~ v^2=v_iv^i=a(r_c)\d_{ij}v^iv^j,~~v^i\equiv
\b^i\sqrt{\frac{h(r_c)}{a(r_c)}}.
\eeq
Thus we  can get the boosted metric
\begin{align}
ds_{p+2}^{2}& = \frac{d\tau^2}{1-v^2/h(r_c)}\(-h(r)+\frac{a(r)}{a(r_c)}v^2 \)
+2\g d\tau dr - \frac{2\g v_i}{h(r_c)}d x^i d r \nn \\
&+
\frac{2v_i}{1-v^2/h(r_c)}\(\frac{h(r)}{h(r_c)}-\frac{a(r)}{a(r_c)}\)d
x^i d\tau \nn
 \\ &+ \left[a(r)\delta_{ij}- \frac{v_i v_j}{h(r_c)(1-v^2/h(r_c))}\(\frac{h(r)}{h(r_c)}
 -\frac{a(r)}{a(r_c)}\)\right]d x^i d x^j.
 \label{eq:metric}
\end{align}
The second is a transformation of $r$ and associated
re-scalings of $\tau$ and $x^i$
\begin{equation} r\rightarrow k(r), \qquad \t \rightarrow \t
\sqrt{\frac{h(r_c)}{h[k(r_c)]}}, ~~~~~x^i \rightarrow x^i
\sqrt{\frac{a(r_c)}{a[k(r_c)]}},\label{t2}
\end{equation}
where we consider the case with $k(r)$ being a linear function of $r$ as $k(r)=br+c$,
with $b$ and $c$  two constants. In this case the general metric (\ref{gm}) becomes
\begin{equation} ds_{p+2}^{2}=-h[k(r)]\frac{h(r_c)}{h[k(r_c)]}d\tau^{2}
+2b\sqrt{\frac{h(r_c)}{h[k(r_c)]}}d\tau
dr+a[k(r)]\frac{a(r_c)}{a[k(r_c)]}dx_{i}dx^{i}.
\end{equation}
When the solution describes a black brane, the cutoff surface $r_c$ is required to be outside
 the horizon $r_c>r_h$. When $h(r)=r,~a(r)=1$, the metric (\ref{gm})  just describes a flat
 space-time written in the ingoing Rindler coordinates, and the
  transformations (\ref{t1}) and (\ref{t2}) agree with those in~\cite{Skenderis} if we take $k(r)=r-r_h$.

After taking the two coordinate transformations one after another,
the resulted metric still solves the corresponding gravitational
field equations. But if we further promote $v_i$ and  $\d k(r)\equiv
k(r)-r=\(b-1\)r+c$ to
 be dependent on the coordinates $x^a$, (that is $v_i, b, c$  are no longer constants), the transformed
 metric is no longer an exact solution of gravitational field equations. In order to solve the gravity equations
 of motion, we take the
  so-called hydrodynamics expansion and non-relativistic limit. Namely we will take the scaling
\begin{equation}
\partial_{r}\sim \epsilon^0, ~~\partial_{i}\sim \epsilon^1,~~
\partial_{\tau}\sim \epsilon^{2}
~~~~~i,j=1,..p\label{eq11}
\end{equation}
together with
\begin{equation}v_{i}\sim \epsilon,~~~~~\d k(r)\sim
\epsilon^{2},
\end{equation}
where $\epsilon$ will be viewed as an expansion parameter.

 As $r$ is arbitrary between $r_h$ and $r_c$, we
demand both $(b-1)$ and $c$ scale as $\epsilon^{2}$. Then up to
order $\epsilon^{2}$, one has
\begin{equation}
 h[k(r)]=h(r)+h'(r)\d k(r),
~~a[k(r)]=a(r)+a'(r)\d k(r),
\end{equation}
and the transformed metric changes to
\begin{align}
ds_{p+2}^{2}=&-h(r)d\tau^{2}+2d\tau dr+a(r)dx_{i}dx^{i}\notag\\
&-2\left(\frac{a(r)}{a(r_{c})}-\frac{h(r)}{h(r_{c})}\right)v_{i}dx^{i}d\tau
-2\frac{v_i}{h(r_{c})}dx^{i}dr \nn
\\
&+\left(\frac{a(r)}{a(r_{c})}-\frac{h(r)}{h(r_{c})}\right)\left[v^{2}d\tau^{2}
+\frac{v_{i}v_{j}}{h(r_{c})}dx^{i}dx^{j}\right]+\frac{v^{2}}{h(r_{c})} d\tau dr \notag  \\
&-h(r)\(\frac{h'(r)\d k(r)}{h(r)}-\frac{h'(r_c)\d
k(r_c)}{h(r_c)}\)d\tau^2+2\(\(b-1\)
-b\frac{h'(r_c)\d k(r_c)}{2h(r_c)}\)d\tau dr\notag\\
&+a(r)\(\frac{a'(r)\d k(r)}{a(r)}-\frac{a'(r_c)\d
k(r_c)}{a(r_c)}\)dx_{i}dx^{i}+{O}(\e^3). \label{gmt}
\end{align}
The first and second lines of metric (\ref{gmt}) are of order
$\epsilon^0$ and $\epsilon^1$ respectively, the other lines  are all
of order $\epsilon^2$.  Note that if one takes
\begin{equation}
 h(r)=r,~~~a(r)=1 ~~~k(r)=r-2P, \label{eq15}
\end{equation}
and the corresponding non-relativistic scaling
\begin{equation}
v_i=v_i^{\e}(\tau,x^i)=\e v_i(\e^2\tau,\e x^i), ~~~P=P^{\e}
(\tau,x^i)=\e^2 P(\e^2\tau,\e x^i),\label{vP}
\end{equation}
it is easy to see that the metric (\ref{gmt}) is the same as the
one in~\cite{Bredberg}, up to order $\epsilon^{2}$.

Next we consider the general asymptotically AdS black brane
solutions. Following \cite{Bhattacharyya:2008jc}, we take the AdS radius to be unit so that
\begin{equation}
h(r)=r^2f(r), ~~~~~a(r)=r^2,\label{eq17}
\end{equation}
where $f(r)$ is an arbitrary function of $r$, but it will be given
in (\ref{bbm}) and (\ref{fgb}) for Einstein gravity and Gauss-Bonnet
gravity respectively. Consider $k(r)$ as a following transformation
\begin{equation}
k(r)=r(1-P) ~\Rightarrow~\d k(r)=-rP,\label{eq18}
\end{equation}
where $P$ is a small parameter. As will be shown shortly, in fact, the parameter $P$ multiplied by a factor
$\frac{r_ch'(r_c)}{2h(r_c)}$ is the pressure density of the dual
fluid.

Substituting (\ref{eq17}) and (\ref{eq18}) in (\ref{gmt}), the first
line of the metric with different $f(r)$ solves the equations of
motion  for the corresponding gravity with a negative cosmological
constant exactly. The remainder terms in (\ref{gmt}) could be
treated as the perturbations of the
 metric, as $(v_i,P)$ will turn out to be the small parameters of the dual non-relativistic fluid.
 To be more specific, we consider $v_i=v_i( x^i,\tau)$ and $P=P ( x^i,
\tau)$ depending on the coordinates $x^a$, but independent of $r$, and take the non-relativistic
hydrodynamic limit in (\ref{vP}), as well as (\ref{eq11}). Then the metric (\ref{gmt}) only solves
 corresponding gravity equations with the cosmological constant $\Lambda=-\frac{p(p+1)}{2}$ at
order $\epsilon^1$. In order to solve the equations to the next
order, we need to add correction terms to the metric
\cite{Skenderis,Bhattacharyya:2008jc}. Let's consider the
constraint equation first \cite{Bhattacharyya:2008jc}. At order $\e^2$, we find that there is only
 one nontrivial constraint
condition: $\p_i v^i=0$. This will turn out to be the
incompressibility of the dual fluid. With this constraint
condition, it turns out that at order $\e^2$, the source terms
only have tensor modes and only the following tensor correction
terms need to be added to the metric:
\begin{equation}
 \frac{r^2}{r_c^2}  F(r)(\p_i v_j+\p_j v_i)dx^{i} dx^{j}
\label{eq19}
\end{equation}
where $F(r)$ is chosen to cancel the source terms at order $\e^2$
and to keep regular at the horizon. In order to keep the induced
metric $\g_{ab}$ invariant, we also need to choose the gauge such
that $F(r_c)=0$ \cite{Skenderis}. Then our final metric up to $\e^2$
is:
\begin{align}
ds_{p+2}^{2}=&-r^2 f(r)d\tau^{2}+2d\tau dr+r^2 dx_{i}dx^{i}\notag\\
&-2\frac{r^2}{r_c^2}\left(1-\frac{f(r)}{f(r_{c})}\right)v_{i}dx^{i}d\tau
-2\frac{v_i}{r_c^2 f(r_{c})}dx^{i}dr \nn
\\
&+\frac{r^2}{r_c^2}\left(1-\frac{f(r)}{f(r_{c})}\right)\left[v^{2}d\tau^{2}
+\frac{v_{i}v_{j}}{r_c^2f(r_{c})}dx^{i}dx^{j}\right]+\frac{v^{2}}{r_c^2f(r_{c})} d\tau dr  \notag\\
&+r^2f(r)\left(\frac{rf'(r)}{f(r)}-\frac{r_cf'(r_c)}{f(r_{c})}\right)Pd\tau^2+\frac{r_cf'(r_c)}{f(r_{c})}Pd\tau dr \notag\\
&+\frac{r^2}{r_c^2} F(r)\(\p_i v_j+\p_j v_i \)dx^{i}dx^{j} +{O}(\e^3), \label{AdSbb}
\end{align}
where the terms in last three lines are all of order $\epsilon^{2}$.



\section{Fluid dual to Einstein Gravity}
\label{sect:ein}
Consider Einstein gravity with a cosmological constant
$\Lambda=-\frac{p(p+1)}{2}$ in $p+2$ dimensions, the equations of motion are given by
\footnote{Here we use $\{\mu ,\nu, \cdots\}$ to stand for the bulk spacetime indices.}
\begin{equation}
E_{\m \n}= R_{\m \n} - \frac{1}{2} g_{\m \n} R-\frac{p(p+1)}{2}\,
g_{\m \n}=0.\label{ein}
\end{equation}
The Einstein's field equations admit the asymptotically $AdS_{p+2}$
black brane solution as
\beq \label{bbm} ds_{p+2}^{2}=-r^2 f(r)d\tau^{2}+2d\tau
dr+r^2dx_idx^i, ~~~~ f(r)=1-\frac{r_h^{p+1}}{r^{p+1}}.
\eeq
Another class of dynamical solutions we are interested here can
also be found in the region between the cutoff surface $\Sigma_c$ and the black brane horizon up to order $\e^2$. The approach to find the solution is described in the previous section, it turns out that the corresponding $F(r)$ term in (\ref{AdSbb}) is
\footnote{We have checked the form $F(r)$ for the case of $0\leq
p\leq 8$ by mathematic calculation, and we expect that this form
 is also valid for higher $p$ \cite{Bhattacharyya:2008mz}.}
\beq
~\label{Fr}~~F(r) =
\int_{r}^{r_c}dx\[\(1-\frac{r_h^{p}}{x^{p}}\)\frac{1}{x^2f(x)}\].
\eeq
Here the boundary condition that $F'(r)$ is regular at $r=r_h$ has
been imposed. Additionally, the integral upper bound has been chosen
to keep $\g_{ab}$ invariant and this matches the result in
\cite{Bhattacharyya:2008kq} when we take the cutoff surface to
infinity. With the constraint condition $\p_i v^i=0$, the metric
(\ref{AdSbb}) solves the Einstein's field equations (\ref{ein}) at
order $\epsilon^{2}$,  i.e., $E_{\m\n}=O(\e^3)$. In what follows,we
will consider the $p=3$ case as a calculation example. The fixed
boundary condition is just the invariant induced metric
$\gamma_{ab}$ on $\Sigma_c$
\beq \g_{ab}dx^a
dx^b=-r_c^2f(r_c)d\tau^2+r_c^2\(dx_1^2+dx_2^2+dx_3^2\). \eeq
With the gravity solution (\ref{AdSbb}), one can calculate the
stress energy tensor of dual fluid. The Brown-York stress energy
tensor $T_{ab}$ evaluated at the cutoff hyper-surface $\Sigma_c$
is~\cite{Strominger}
\beq {T_{ab}={1 \over 8\pi G}
\left(\gamma_{ab}K-K_{ab}+C\gamma_{ab} \right) ,} \eeq
where $K_{ab}$ is the extrinsic curvature tensor of $\Sigma_c$, $K$
is its trace, and $C$ is an ambiguous constant. It is obvious that
$T_{ab}^{C}=\frac{C}{8\pi G}\g_{ab}$ only have order $\e^0$ terms.
After some straightforward calculations, we obtain the stress energy
tensor of the dual fluid as ~\footnote{Here we use the units $16\pi
G=1$ and $f_c=f(r_c),f'_c=f'(r_c),F'_c=F'(r_c), \cdots $ for short.}
\beq
T_{ab}=T^{(0)}_{ab}+T^{(1)}_{ab}+T^{(2)}_{ab}+{O}(\e^3) \eeq
where
\beq \left\{
\begin{aligned}
&T^{(0)}_{ab}dx^adx^b=-r_c^2 f_c\(6\sqrt{f_c}+2C\)d\tau^2+
r_c^2\(6\sqrt{f_c}+2C+\frac{r_c f'_c}{\sqrt{f_c}}\)dx_idx^i \\
&T^{(1)}_{ab}dx^adx^b=-2\frac{r_cf'_c}{\sqrt{f_c}}v_idx^i d\tau\\
&T^{(2)}_{ab}dx^adx^b=\frac{r_cf'_c}{\sqrt{f_c}}\(3r_c^2f_cP+v^2\)
d\t^2+\frac{r_cf'_c}{\sqrt{f_c}}\[\frac{v_iv_j}{r_c^2f_c} +\(1+\frac{r_cf'_c}{2f_c}\)r_c^2P\d_{ij}\]dx^idx^j \\
&\qquad\qquad\qquad-\frac{1+r_c^2f_cF'_c}{r_c\sqrt{f_c}}(\p_iv_j+\p_jv_i)dx^idx^j.
\label{bbst}
\end{aligned}
 \right.
\eeq
Since the conservation equations of the Brown-York stress energy
tensor are just the Gauss-Codazzi formulas of Einstein's field
equations~\cite{Brown:1992br,Strominger}. This means that the
conservation equations of the dual fluid stress energy tensor are
the corresponding constraint equations of gravity equations. With
the conservation equations of the stress energy tensor, we have the
first nontrivial equation at order $\e^2$,
\beq
\p^aT_{a\tau}=\frac{r_cf'_c}{\sqrt{f_c}}\p^iv_i=\frac{4r^4_h}{r^2_c
\sqrt{r^4_c-r^4_h}}\p_iv^i=0 \Rightarrow\p_iv^i=0 .\label{bbc1}
\eeq
This agrees with the constraint equation in the gravity side and
it turns out that the dual fluid is incompressible. Taking this to be the case, to the next order, we have
\beq
\p^a T_{ai}=\frac{f'_c}{r_cf_c\sqrt{f_c}}\[\p_\tau v_i-\frac{f_c\(1+r_c^2f_cF'_c\)}
{f'_c} \p^2v_i+r_c^2f_c\(1+\frac{r_cf'_c}{2f_c}\)\p_iP+v^j\p_j v_i\] =0.\label{bbc2}
\eeq
This would be the constraint equation of the gravity metric of order
$\e^3$ and more correction terms need to be added to solve the
equations of motion of gravity at this order.

In the infinity boundary limit $r_c\rightarrow\infty$, as the cutoff surface $\Sigma_c$ is intrinsic flat, from the surface counter-term approach~\cite{Balasubramanian:1999re}, we have to take $C=-3$ for our $AdS_5$ case in order to remove the divergence in the stress energy tensor. In AdS/CFT, the background metric $h_{ab}$ for the fluid stress energy tensor $\langle T_{ab}\rangle$ is redefined by stripping off the divergent conformal factor from the boundary metric~\cite{Myers:1999psa}
\beq h_{ab}=\lim_{r_c\rightarrow\infty}
{\frac{\gamma_{ab}}{r_c^2}}, \label{bbdt}\qquad
\sqrt{-h}h^{ab}\langle T_{bc}\rangle=
\lim_{r_c\rightarrow\infty}\sqrt{-\g}\g^{ab}T_{bc}. \eeq
Our result can also recover the result for dual fluid on the boundary in the non-relativistic limit~\cite{Bhattacharyya:2008jc,Bhattacharyya:2008kq}, we will see this later. To see clearly the properties of the dual fluid, it is more convenient to rewrite (\ref{bbst}) in the $(\tc,\x^i)$ coordinates:
\beq \left\{
\begin{aligned}
&\T^{(0)}_{ab}d\x^ad\x^b=-\(6\sqrt{f_c}+2C\)d\tc^2+\(6\sqrt{f_c}+2C+\frac{r_c f'_c}{\sqrt{f_c}}\)d\x_{i}d\x^i\\
&\T^{(1)}_{ab}d\x^ad\x^b=-2\frac{r_cf'_c}{\sqrt{f_c}}\b_i d\x^i d\tc \label{bbstc}\\
&\T^{(2)}_{ab}d\x^ad\x^b=\frac{r_cf'_c}{\sqrt{f_c}}\(3P+\b^2\)d\tc^2+\frac{r_cf'_c}{\sqrt{f_c}}\[\b_i\b_j+\(1+\frac{r_cf'_c}{2f_c}\)P\d_{ij}\]d\x^id\x^j\\
&\qquad\qquad\qquad-\(1+r_c^2f_cF'_c\)(\pa_i\b_j+\pa_j\b_i)d\x^id\x^j .\\
\end{aligned}
 \right.
\eeq
In general, the stress energy tensor of relativistic fluid in
$4$-dimensional Minkowski background $\tilde{\g}_{ab}=\eta_{ab}$ can be written as~\cite{Bhattacharyya:2008kq,Landau}
\beq \T_{ab}=\tilde{\r}
\u_a\u_b+\tilde{p}\h_{ab}-2\eta\si_{ab}-\zeta\tilde{\c} \tilde h_{ab},
\label{eq32}
\eeq
where
\beq \h_{ab}=\u_a\u_b+\tilde{\g}_{ab},~~~
\si_{ab}=\frac{1}{2}\h_{ad}\h_{be}(\pa^d\u^e+\pa^e\u^d)-\frac{1}{3}\tilde{\c}\h_{ab},
~~~\tilde{\c}=\pa_a\u^a.
\label{eq33}
\eeq
In the non-relativistic limit and with the incompressible condition, up to order $\e^2$, we have
\beq
\T_{\t\t}=\tilde{\r}+(\tilde{\r}+\tilde{p})\b^2,~~~ \T_{\t i}=-(\tilde{\r}+\tilde{p})\b_i,~~~
\T_{ij}=(\tilde{\r}+\tilde{p})\b_i\b_j+\tilde{p}\d_{ij}-\eta(\pa_i\b_j+\pa_j\b_i).
\label{eq34}
\eeq
Comparing our dual fluid stress energy tensor (\ref{bbstc}) with
this form, we can get from $\T_{ab}^{(0)}$ in (\ref{bbstc}) the
energy density and pressure of dual fluid at order $\epsilon^0$
as
\beq
\label{bbbg}\r_0=-(6\sqrt{f_c}+2C),~~p_0=6\sqrt{f_c}+2C+\frac{r_cf'_c}{\sqrt{f_c}},
~~\o_0\equiv\r_0+p_0=\frac{r_cf'_c}{\sqrt{f_c}}.
 \eeq
  Further we can obtain  from
$\T_{ab}^{(2)}$ in (\ref{bbstc})  the $r_c$ dependent dynamic
viscosity \beq
\label{bbeta}\eta_c\equiv\eta(r_c)=\(1+r_c^2f_cF'_c\)=\frac{r_h^3}{r_c^3}=\frac{1}{16\pi
G}\frac{r_h^3}{r_c^3}. \eeq
As $\p_i\b_j\sim\e^2$, we see the viscosity $\eta_c$ is of order
$\e^0$, it is better to compare this with the background of the fluid entropy density, which is just
the background black brane horizon entropy density. The entropy density $s_c$ associated with the
cutoff surface is~\cite{Strominger}
\beq \label{bbs}s_c\equiv
s(r_c)=\frac{1}{4G}\frac{r_h^3}{r_c^3}
~~~\Longrightarrow~~~\frac{\eta_c}{s_c}=\frac{1}{4\pi}.
\eeq
We see that the ratio of shear viscosity to entropy density is
independent of the cutoff $r_c$, which means that the ratio does not
run with the cutoff. Note that for the static background solution
(\ref{bbm}), the Hawking temperature $T_H$ of the horizon and the
local temperature $T_c$  on the cutoff surface respectively are
\beq
\label{bbT}T_H=\frac{\[r^2f(r)\]'}{4\pi}|_{r=r_h}=\frac{r_h}{\pi},
~~~~T_c=\frac{T_H}{\sqrt{r_c^2f_c}}=\frac{1}{\sqrt{r_c^2f_c}}\frac{r_h}{\pi}.
\eeq
We see that the following thermodynamic relation still holds
\beq \label{bbo}\o_0=T_cs_c=\frac{r_c
f'_c}{\sqrt{f_c}}=\frac{4r_h^4}{r_c^2\sqrt{r_c^4-r_h^4}}, \eeq for
the dual fluid on the cutoff surface. In addition, the dimensionless
coordinate invariant diffusivity $\bar{D}_c$ defined in
\cite{Strominger} is found to be
\beq \label{bbd}\bar{D}_c\equiv
T_c\frac{\eta_c}{\o_0}=\frac{\eta_c}{s_c}=\frac{1}{4\pi}, \eeq
a universal constant.

Next let's read off in the $(\tc,\x^i)$ coordinates the energy  density and pressure of the dual fluid,
to order $\e^2$,
\begin{align}
&\r_c=\r_0+3\o_0P, \label{eq41}
~~~~p_c=p_0+\o_0\(1+\frac{r_cf'_c}{2f_c}\)P.
\end{align}
The Hamiltonian constraint on $\Sigma_c$ would play a role analogous to that of the equation of
state for a conventional fluid~\cite{Skenderis}. For the fluid dual to Einstein gravity with
a negative cosmology constant $\Lambda$, if we define $\hat{T}_{ab}=\T_{ab}-2C\g_{ab}=2(\g_{ab}K-K_{ab})$,
 the Hamiltonian constraint would turn out to be $\hat{T}^2-p\hat{T}_{ab}\hat{T}^{ab}+8p\Lambda=0$. This
 constraint provides the relation between the energy density and pressure of the dual fluid. If we
 evaluate the Hamiltonian constraint using the Brown-York stress tensor (\ref{bbstc}), the first
 nontrivial equation is encountered at order $\e^2$, which is just the
incompressible
 condition $\tilde{\p}_i \b^i=0$. Taking this to be the case and substituting (\ref{eq34}) into the
  constraint, we can get $(\tilde{\rho}+2C)(\tilde{\rho}+3\tilde{p}-4C)+72=0$ up to order $\e^2$. It
  can be checked that the energy density and pressure read off from the Brown-York tensor of the dual
  fluid satisfy this equation of state, by use of $(\r_c, p_c)$ in
  (\ref{eq41}).

We can also calculate the trace of the stress energy tensor
(\ref{bbstc}), to order $\e^2$,
\beq
\T_c\equiv\T_{ab}\tilde{\g}^{ab}=4\[3\(f_c^{1/2}+f_c^{-1/2}\)+2C\]+\frac{3\o_0^2}{2\sqrt{f_c}}P,
\eeq
which in general does not vanish. However, if take the cutoff
surface $r_c\rightarrow\infty$, and consider the conformal
invariance of dual field theory on the AdS boundary, namely,
$\T_c\rightarrow0$, we can also recover the counterterm factor
$C=-3$. In addition, we find that
\beq
\lim_{r_c\rightarrow\infty}\r_cr_c^4=3r_h^4\(1+4P\),\qquad
\lim_{r_c\rightarrow\infty}p_cr_c^4=r_h^4\(1+4P\).\label{brp}
\eeq
Note that $r_c^4$ in (\ref{brp}) should be absorbed by the dual fluid stress energy
 tensor $\langle T_{ab}\rangle$ defined in the flat spacetime $h_{ab}$ in (\ref{bbdt}), the results reproduce
the holographic values.

Following~\cite{Bhattacharyya:2008kq}, we define the pressure
density as
\beq
P_c\equiv\frac{p_c-p_0}{\r_0+p_0}=\(1+\frac{r_cf'_c}{2f_c}\)P
\label{pc}\eeq
and the kinematic viscosity as
\beq
\nu_c\equiv\frac{\eta_c}{\o_0}=\(\frac{r_h}{r_c}\)^3\frac{\sqrt{f_c}}{4}\(\frac{r_c}{r_h}\)^4=\frac{1}{4\pi
T_c}\label{nu}.
\eeq
Then we can rewrite the conservation equation (\ref{bbc2}) in the
$(\tc, \x^i)$  coordinates with Minkowski metric $\eta_{ab}$ and
3-velocity $\b_i$ as
\beq
\frac{1}{\o_0}\pa^a \label{bns} \T_{ai}=\pa_\t\b_i-\n_c\pa^2\b_i+\pa_iP_c+\b^j\pa_j\b_i =0.
\eeq
This is nothing, but precisely the incompressible Navier-Stokes
equation up to order $\e^3$ of the dual fluid in flat spacetime.


\section{Fluid dual to Gauss-Bonnet Gravity}
\label{sect:gb}

The Einstein-Hilbert action with the Gauss-Bonnet term
$\mathcal{L}_{GB}=\(R^2-4R_{\mu \nu}R^{\mu \nu}+R_{\mu \nu
\sigma\tau}R^{\mu \nu \sigma \tau}\)$, and a negative cosmological
constant $\Lambda=-\frac{p(p+1)}{2\ell^2}$ in $p+2$ dimensions can
be written as
\begin{eqnarray}
\label{action} S=\frac{1}{16 \pi G}\int_\mathcal{M}~d^{p+2}x
\sqrt{-g} \left(R-2 \Lambda+\alpha\mathcal {L}_{GB}\right),
\end{eqnarray}
where $\alpha$ is the Gauss-Bonnet coefficient with the same
dimension as square of AdS radius $\ell$. As $\mathcal{L}_{GB}$ is nontrivial for $p\geq3$,
we will consider the $p=3$ case and take the unit $AdS$ radius $\ell=1$ in what follows. With the
corresponding surface terms~\cite{Myers:1987yn,Brihaye:2008xu}, we can obtain the equations of
 motion  for the Gauss-Bonnet gravity by varying the action (\ref{action}) with respect to metric
\begin{eqnarray}
\left\{
\begin{aligned}
\label{GBeqs} &R_{\mu \nu } -\frac{1}{2}Rg_{\mu \nu}-6g_{\mu \nu
}+\alpha H_{\mu \nu}=0~,\\
\label{Hmn}
&H_{\mu \nu}=2(R_{\mu \sigma \kappa \tau }R_{\nu }^{\phantom{\nu}%
\sigma \kappa \tau }-2R_{\mu \rho \nu \sigma }R^{\rho \sigma
}-2R_{\mu \sigma }R_{\phantom{\sigma}\nu }^{\sigma }+RR_{\mu \nu
})-\frac{1}{2}g_{\mu \nu }\mathcal {L}_{GB}.
\end{aligned}
 \right.
\end{eqnarray}
The Gauss-Bonnet gravity with a negative cosmological constant is
solved by \cite{Deser,Cai:2001dz} with spherical symmetry. Here we
consider the $5$-dimensional black brane solution written in the
Eddington-Finkelstin coordinates~\cite{Cai:2001dz,Hu:2010sn}
\begin{equation}
\left\{
\begin{aligned}
&ds^2 = - r^2 f(r)d\t^2 + 2 d\t dr + r^2(dx_1^2 +dx_2^2 +dx_3^2)\\ &f(r)
= \frac{1}{4\alpha}\(1-\sqrt{1-8\alpha\(1-\frac{r_h^4}{r^{4}}\)}\).
\label{fgb}
\end{aligned}
 \right.
\end{equation}
Note that when $r\rightarrow\infty$,
$f(r)\rightarrow\frac{1}{4\a}\(1-\sqrt{1-8\a}\)$. One can define the
effective $AdS$ radius $\ell_e$ in the Gauss-Bonnet gravity as
\begin{eqnarray}
\label{lc}
\ell_e^2=\frac{1+\sqrt{1-8\alpha}}{2}=\frac{4\a}{1-\sqrt{1-8\alpha}}.
\end{eqnarray}
We find that in the case of the Gauss-Bonnet gravity, the correction term (\ref{eq19}) and the
solution (\ref{AdSbb}) also work with $p=3$. But the form $F(r)$ has to be replaced by
\begin{eqnarray}
F(r)=\int_{r}^{r_c}dx\[1-\frac{r_h^3}{x^3}\frac{1-8\a}{\(1-2\a\[2f(x)+xf'(x)\]\)}\]\frac{1}{x^{2}f(x)}.
\end{eqnarray}
With the constraint equations $\p_iv^i=0$, we have checked that
(\ref{AdSbb}) solves the equations of motion of the Gauss-Bonnet
gravity equations (\ref{GBeqs}) up to order $\e^2$. Here when $r_c
\to \infty$, the form $F(r)$ gives the result in~\cite{Hu:2010sn}.

From~\cite{Davis:2002gn,Brihaye:2008xu}, we can take the Brown-York stress energy tensor for the Gauss-Bonnet
gravity as
\begin{equation}
 T_{ab}=\frac{1}{8 \pi G}\[K\gamma _{ab}-K_{ab}-2\alpha \(3J_{ab}-J\gamma_{ab}+2\hat{P}_{acdb}K^{cd}\)
 +C\gamma _{ab}\], \label{TabCFT}
\end{equation}
where
\beq \hat{P}_{abcd} = \R_{abcd} + 2 \R_{b[c} \g_{d]a} - 2 \R_{a[c}
\g_{d]b} + \R \g_{a[c}\g_{d]b} \label{Pdef}
\eeq
is an intrinsic tensor associated with the flat induced metric $\gamma_{ab}$. So in our case $\hat{P}_{abcd}$
 will not make any contribution to the stress energy tensor of the dual fluid. Namely, we have
\begin{equation}
 T_{ab}=\frac{1}{8 \pi G}\[K\gamma _{ab}-K_{ab}-
 2\alpha \(3J_{ab}-J\gamma_{ab}\)+C\gamma _{ab}\], \label{GBTab}
\end{equation}
with
\begin{equation}
J_{ab}=\frac{1}{3}%
(2KK_{ac}K_{b}^{c}+K_{cd}K^{cd}K_{ab}-2K_{ac}K^{cd}K_{db}-K^{2}K_{ab})~.
\label{Jab}
\end{equation}
By use of Gauss-Codazzi equations, the conservation of stress energy
tensor (\ref{GBTab}) can also be deduced from the equations
(\ref{GBeqs}) of motion of the Gauss-Bonnet
gravity~\cite{Davis:2002gn}.  Through some straightforward
calculations, the stress energy tensor of the dual fluid to the
Gauss-Bonnet gravity is found to be
\beq
T_{ab}=T^{(0)}_{ab}+T^{(1)}_{ab}+T^{(2)}_{ab}+{O}(\e^3) \eeq
where
\beq
\left\{
\begin{aligned}
  \label{gbst}
&T^{(0)}_{ab}dx^adx^b=\(-6\sqrt{f_c}-2C+8\a f_c\sqrt{f_c}\)r_c^2 f_cd\tau^2\\
&\qquad\qquad\qquad\(6\sqrt{f_c}+2C-8\a f_c\sqrt{f_c}+\(1-4 \a f_c\)\frac{r_c f'_c}{\sqrt{f_c}}\)r_c^2dx_idx^i\\
&T^{(1)}_{ab}dx^adx^b=-2\(1-4\a f_c\)\frac{r_cf'_c}{\sqrt{f_c}}v_i dx^i d\tau \\
&T^{(2)}_{ab}dx^adx^b=\(1-4\a f_c\)\(3r_c^2f_cP+v^2\)\frac{r_cf'_c}{\sqrt{f_c}}d\t^2\\
&\qquad\qquad\qquad+\(1-4 \a f_c\)\frac{r_cf'_c}{\sqrt{f_c}}\[\frac{v_iv_j}{r_c^2f_c} +\(1+\frac{r_cf'_c}{2f_c}\)r_c^2P\d_{ij}\]dx^idx^j \\
&\qquad\qquad\qquad-\[1-2 \a \(2f_c+r_cf'_c\)\] \frac{1+r_c^2f_cF'_c}{r_c\sqrt{f_c}}(\p_iv_j+\p_jv_i)dx^idx^j\\
\end{aligned}
  \right.
\eeq
The corresponding conservation equations for the stress energy
tensor at order $\e^2$ are
\beq
 \p^a T_{a\tau}=\frac{\(1-4\a f_c\)r_cf'_c}{\sqrt{f_c}}\p^iv_i
=\frac{4}{\sqrt{f_c}}\frac{r_h^4}{r_c^4}\p_iv^i=0 ~\Rightarrow ~
\p_iv^i=0, \eeq
which gives the incompressible condition again, and  to the next
order
\beq \p^a T_{ai}=\frac{\(1-4\a
f_c\)f'_c}{r_cf_c\sqrt{f_c}}\[\p_\tau v_i-\frac{\eta_c}{\(1-4\a
f_c\)}\frac{f_c}{f'_c}  \label{gbc2}
\p^2v_i+r_c^2f_c\(1+\frac{r_cf'_c}{2f_c}\)\p_iP+v^j\p_j v_i\] =0,
\eeq
where $\eta_c=\[1-2 \a \(2f_c+r_cf'_c\)\] (1+r_c^2f_cF'_c)$,
which will turn out to be the fluid's dynamic viscosity. The
equation (\ref{gbc2}) gives the constraint condition of the
solutions of the Gauss-Bonnet gravity (\ref{GBeqs}) at order $\e^3$.

In the $(\tc,\x^i)$ coordinates, the corresponding stress energy tensor of the dual fluid is
\beq
\left\{
\begin{aligned}
  \label{gbstc}
&\T^{(0)}_{ab}d\x^ad\x^b=\(-6\sqrt{f_c}-2C+8\a f_c\sqrt{f_c}\)d\tc^2\\
&\qquad\qquad\qquad\(6\sqrt{f_c}+2C-8\a f_c\sqrt{f_c}+\(1-4 \a f_c\)\frac{r_c f'_c}{\sqrt{f_c}}\)d\x_{i}d\x^i\\
&\T^{(1)}_{ab}d\x^ad\x^b=-2\(1-4\a f_c\)\frac{r_cf'_c}{\sqrt{f_c}}\b_i d\x^i d\tc \\
&\T^{(2)}_{ab}d\x^ad\x^b=\(1-4\a f_c\)\(3P+\b^2\)\frac{r_cf'_c}{\sqrt{f_c}}d\tc^2\\
&\qquad\qquad\qquad+\(1-4\a f_c\)\frac{r_cf'_c}{\sqrt{f_c}}\[\b_i\b_j+\(1+\frac{r_cf'_c}{2f_c}\)P\d_{ij}\]d\x^id\x^j\\
&\qquad\qquad\qquad-\[1-2 \a \(2f_c+r_cf'_c\)\] \(1+r_c^2f_cF'_c\)\(\pa_i\b_j+\pa_j\b_i\)d\x^id\x^j,\\
\end{aligned}
  \right.
\eeq
from which we can obtain the energy density and pressure of the dual fluid at order $\epsilon^0$
 \beq \left\{
\begin{aligned}
  \label{gbrp}
&\r_0=-6\sqrt{f_c}-2C+8\a f_c\sqrt{f_c},\\
&p_0=6\sqrt{f_c}+2C-8\a f_c\sqrt{f_c}+\(1-4 \a f_c\)\frac{r_c f'_c}{\sqrt{f_c}}\\
&\o_0\equiv\r_0+p_0=\(1-4 \a
f_c\)\frac{r_cf'_c}{\sqrt{f_c}}=\frac{4}{\sqrt{f_c}}\frac{r_h^4}{r_c^4},
\end{aligned}
  \right.
\eeq
while to order $\e^2$, they are
 \beq \r_c=\r_0+3\o_0P,\qquad
p_c=p_0+\o_0\(1+\frac{r_cf'_c}{2f_c}\)P.
 \eeq
And the trace of the stress tensor (\ref{gbstc}) is
\beq
\T_c\equiv\T_{ab}\tilde{\g}^{ab}=4\[3\(f_c^{1/2}+f_c^{-1/2}\)-2\a f_c^{-3/2}+2C\]+\frac{3\o_0^2}
{2\sqrt{f_c}\(1-4\a f_c\)}P~.
\eeq
If one takes the limit $r_c\rightarrow\infty$, some surface counter-terms \cite{Brihaye:2008xu} are needed.
Since the intrinsic Riemann tensor associated with $\g_{ab}$ vanishes, in our case, we need only to
 take $C=-\frac{2+\sqrt{1-8\a}}{\ell_e}$, which is also consistent with a vanishing trace $\T_c =0$ when
$r_c\rightarrow\infty$. Thus we have
\beq
\lim_{r_c\rightarrow\infty}\r_c r_c^4=3\ell_er_h^4\(1+4P\),\qquad
\lim_{r_c\rightarrow\infty}p_c r_c^4=\ell_er_h^4\(1+4P\).\label{gbrp}
\eeq
Once again, transforming back to the $(\t,x^i)$ coordinates in the boundary metric $h_{ab}$ (\ref{bbdt}),
 the divergent factor
$r_c^4$ could be absorbed. We have checked the result and shown that they match with the stress energy
tensor of the dual fluid on the boundary~\cite{Hu:2010sn}.

In this case, we find that the dynamic viscosity
\beq \eta_c=\[1-2\a \(2f_c+r_cf'_c\)\] (1+r_c^2f_cF'_c)=\frac{1}{16\pi
G}\frac{r_h^3}{r_c^3}\(1-8\a\). \eeq
The black brane entropy density associated with the cutoff surface is given by
\beq s_c =\frac{1}{4
G}\frac{r_h^3}{r_c^3}~~\Longrightarrow~~~\frac{\eta_c}{s_c}=\frac{1}{4\pi}\(1-8\a\).
\eeq
This ratio of shear viscosity to entropy density is the same as that of the dual fluid to the Gauss-Bonnet
 gravity on the
boundary~\cite{Brigante:2007nu,{Brigante:2008gz},{KP},{Cai:2009zv}}.
Thus, through the hydrodynamic expansion method, we find that the ratio  for the dual fluid on the
cutoff surface is  independent of the cutoff, and does not run with the cutoff~\cite{Hong.Membrane}.
In addition, we find the thermodynamic relation holds and the diffusivity is changed as
\beq
\o_0=T_cs_c,~~~~~~~~~~\bar{D}_c=\frac{\eta_c}{\o_0}T_c=\frac{1}{4\pi}\(1-8\a\),
\eeq
where the local temperature on the cutoff surface is
\beq
~~T_c=\frac{T_H}{\sqrt{r_c^2f_c}}=\frac{1}{\sqrt{r_c^2f_c}}\frac{\[r^2f(r)\]'}{4\pi}|_{r=r_h}=\frac{1}
{\sqrt{r_c^2f_c}}\frac{r_h}{\pi}. \eeq
Finally we give the incompressible Navior-Stokes equations dual to the Gauss-Bonnet gravity
\beq
  \frac{1}{\o_0}\pa^a \T_{ai}=\pa_\tau
\b_i-\n_c\pa^2\b_i+\pa_iP_c+\beta^j\pa_j\b_i =0. \eeq
where
\beq \nu_c=\frac{\eta_c}{\o_0}=\frac{\(1-8\a\)}{4\pi T_c},
~~~~~P_c=\(1+\frac{r_cf'_c}{2f_c}\)P. \eeq


\section{Conclusions}
\label{sect:co}

By use of the non-relativistic hydrodynamic expansion method, we have solved the equations of motion for
 Einstein gravity and
Gauss-Bonnet gravity with a negative cosmological constant,
respectively, and obtained black brane metrics in the region between a finite cutoff surface  $\Sigma_c$ and
 the black brane horizon, up to order $\epsilon^2$. We have calculated the stress energy tensor of the dual
  fluid on the cutoff surface through the Brown-York tensor on the surface~\cite{Strominger}. And then we
  have discussed some properties of the dual fluid.  It turns out that the dual fluid is an incompressible one
   in both cases, obeys the Navier-Stokes equations, but with different kinematic viscosity. In both cases, the
    ratio of shear viscosity to entropy density is independent of the cutoff, it means that the ratio does not
     run with the cutoff surface. When one moves the cutoff surface to spatial infinity, namely takes the
      limit $r_c \to \infty$, our results can recover those for dual fluids on the boundary
      of AdS space~\cite{Bhattacharyya:2008jc}. In the near horizon limit, our results should also be in
      contact with the membrane paradigm in \cite{Eling:2009pb}, where the horizon dynamics is described
      by incompressible Navier-Stokes equations in the non-relativistic hydrodynamic limit.

 Our results presented in this paper together with those
  in~\cite{Strominger,Bredberg,Skenderis} show that the fluids on a finite cutoff surface always obey
   the incompressible Navier-Stokes equations, for dual gravity solutions up to second order of the non-relativistic
    hydrodynamic expansion parameter. This may reveal some insights for the holographic dual to
    asymptotically flat spacetimes.  The study of holography on a finite cutoff may also be helpful
 to understand the microscopic origin of gravity.  It is shown in \cite{Strominger} that the entropy
 flow equation along the radial coordinate is equivalent to a radial  Einstein equation.
 We have checked that this also holds in the Gauss-Bonnet  gravity.

Finally we would like to stress that the black brane solutions
obtained in this paper are up to second order of the
non-relativistic hydrodynamic expansion parameter $\e$. In
principle, following \cite{Skenderis}, we can obtain corresponding
black brane solutions to an arbitrary order of the expansion
parameter.  With the resulting solutions, by calculating the
corresponding Brown-York tensor in the gravity side, we can obtain
the stress energy tensor of dual fluids on the cutoff surface, and
then obtain other transport coefficients of the dual fluids.

\section*{Acknowledgements}

This work was supported in part by the National Natural Science
Foundation of China (No.10821504, No.10975168 and No.11035008), and
in part by the Ministry of Science and Technology of China under
Grant No. 2010CB833004. LL and YLZ are extremely grateful to Li-Ming
Cao, Ya-Peng Hu and Zhang-Yu Nie for calculations by Mathematica, as
well as Bin Hu, Hai-Qing Zhang, Huai-Fan Li, Song He, Wei Gu for
useful discussions. Finally we would like to thank the two referees
for their helpful comments and suggestions.

\appendix

\end{document}